\documentclass[prd,twocolumn,floats,floatfix,nofootinbib]{revtex4}
\usepackage[dvips]{graphicx}
\usepackage{graphics}
\usepackage{dcolumn}
\usepackage{amssymb}
\usepackage{bm}
\usepackage{amsmath}
\usepackage{color}

\def\spose#1{\hbox to 0pt{#1\hss}}

\def\lta{\mathrel{\spose{\lower 3pt\hbox{$\mathchar"218$}}
     \raise 2.0pt\hbox{$\mathchar"13C$}}}
\def\gta{\mathrel{\spose{\lower 3pt\hbox{$\mathchar"218$}}
     \raise 2.0pt\hbox{$\mathchar"13E$}}}
\newcommand{\be}{\begin{equation}}
\newcommand{\en}{\end{equation}}
\newcommand{\bea}{\begin{eqnarray}}
\newcommand{\ena}{\end{eqnarray}}

\begin{document}

\title{Boosted Kerr black holes in general relativity}

\author{Ivano Dami\~ao Soares$ $}

\affiliation{
$ $Centro Brasileiro de Pesquisas F\'{\i}sicas -- CBPF/MCTI, \\ Rua Dr. Xavier Sigaud, 150, Urca,
CEP 22290-180, Rio de Janeiro, Brazil,\\ email:ivano@cbpf.br}

\date{\today}

\begin{abstract}

A solution of Einstein's vacuum field equation is derived that describes
a general boosted Kerr black hole relative to a Lorentz frame at future null infinity.
The metric contains five independent parameters --- mass $m$, rotation $\omega$,
boost parameter $v/c$ and the boost direction defined by $(n_1,n_2,n_3)$
satisfying $(n_1)^2+(n_2)^2+(n_3)^2=1$ --- and reduces to the Kerr black hole when
the boost parameter is zero and $n_1=1$. The solution describes the most general configuration that
an astrophysical black hole must have. The black hole rotates about the $z$-axis with angular momentum
proportional to $m \omega$ and the geometry has just one Killing vector $\partial/\partial{u}$,
where $u$ is the retarded time coordinate.
The boost turns the ergosphere asymmetric, with dominant lobes in the direction opposite to the boost.
The event and Cauchy horizons, defined for the case $\omega < m$, are specified respectively by the
radii $r_{\pm}=m \pm \sqrt{m^2-\omega^2}$. The horizons are topologically  spherical and
the singularity has the topology of a circle on planes that are orthogonal to the boost direction.
We argue that this black hole geometry is the natural set to describe the remnants of the recently
observed gravitational wave events $GW150914$, $GW151226$, $GW170814$ and $GW170817$\cite{gw1,gw2,gw3,gw4}.
In the conclusions we discuss possible astrophysical processes in the asymmetric ergosphere
and the electromagnetic dynamical effects that may result from the rotating black hole
moving at relativistic speeds.

PACS numbers:

\end{abstract}

\maketitle

\section{Introduction\label{sectionI}}

The Kerr black hole\cite{kerr0} is an exact stationary solution of vacuum Einstein's equations
of general relativity that describes a rotating black hole with two parameters, mass and angular momentum,
and has the Schwarzschild black hole\cite{schwarz} as its static configuration limit.
The Kerr solution turned out to be of fundamental importance to the understanding of astrophysical
processes involved in objects with a tremendous output of energy as quasars, pulsars
and active galactic nuclei (AGNs).
The recent direct observations of the gravitational wave emission from binary black hole mergers --
GW150914\cite{gw1}, GW151226\cite{gw2}, GW170814\cite{gw3}, GW170817\cite{gw4} -- by the
the LIGO Scientific Collaboration and the Virgo Collaboration showed that the initial black holes
of each binary had mass ratios  ~$\alpha_{GW150914} \simeq 0.8$, ~$\alpha_{GW151226} \simeq 0.53$,
~$\alpha_{GW170104} \simeq 0.62$ and ~$\alpha_{GW170914} \simeq 0.83$, respectively. The nonequal mass
of the initial black holes in the observed binaries imply that the gravitational waves emitted have a non-zero gravitational
wave momentum flux, indicating that the remnant black hole is a Kerr black hole boosted
along a particular direction relative to the asymptotic Lorentz frame at null infinity where such
emissions have been detected. In this sense the remnant black hole description must contain
additional parameters -- the boost parameters -- connected to
its motion relative to the observation frame.
The boost of the remnant black hole results from the net momentum flux of the gravitational waves emitted
in the collision and merger of two nonequal mass black holes that generated the remnant.

\par
The main object of this paper is to describe an exact solution of a general boosted Kerr black hole relative
to an asymptotic Lorentz frame at future null infinity. This solution corresponds to the most general
configuration that an astrophysical remnant black hole must have, in particular as the remnant configuration
of the collision and merger of black holes recently observed in the direct detection of gravitational
waves\cite{gw1,gw2,gw3,gw4}.
The derivation and interpretation of this solution will be framed in the Bondi-Sachs (BS) characteristic
formulation of gravitational wave emission in general relativity\cite{bondi1,sachs, sachs1, sachs2},
where we have a clear and complete derivation of physical quantities and its conservation laws,
connected to the radiative wave transfer of energy and momentum, namely,
the mass and momentum extracted of the system by the gravitational waves emitted,
evaluated at the future null infinity ($r \rightarrow \infty$), where the spacetime is asymptotically flat.
In the BS formulation the conservation laws and final values of the conserved quantities
are exact. The formulation\cite{bondi1,sachs} relies on (i) expanding the metric functions
in a power series of $1/r$ (where $r$ is the luminosity distance), (ii) taking into in account
the BS boundary conditions (connected to the asymptotic flatness of the spacetime and the outgoing wave condition),
(iii) using Einstein's vacuum equations and (iv) eliminating some arbitrary functions that arise in the
integration scheme. This procedure is far from being trivial, and furthermore the coordinate system used
may present singularities for $r$ sufficiently small\cite{bondi1,sachs,sachs1,sachs2}. In this sense the use of the above procedures
to obtain the exact metric of the remnant spacetime, expected to be that of a general boosted Kerr
black hole, remains yet to be done. An analogous difficulty occurs in the case of 1+3 Numerical
Relativity simulations.
\par
To circumvent this problem we will undertake the integration of Einstein's equation for
stationary twisting Petrov D vacuum spacetimes, as discussed in the present paper.
Among this Petrov type D class we will obtain a spacetime solution that corresponds
to a Kerr black hole with additional parameters connected
to the motion of the black hole relative to an asymptotic Lorentz frame at future null infinity;
as the boost parameters go to zero we recover the Kerr metric\cite{kerr0}.
This metric is therefore a candidate to describe an astrophysical remnant black hole,
in particular the remnant configuration of the collision and merger of black holes
recently observed in the direct detection of gravitational waves\cite{gw1,gw2,gw3,gw4}.
We also discuss as the boosted black hole solution can be a natural set for astrophysical processes
connected to the asymmetry of the ergosphere and to electromagnetic dynamical effects that result from
the rotating black hole moving at relativistic speeds in a direction not coinciding with
the rotation axis of the black hole. These effects may correspond to
the electromagnetic counterpart of the gravitational wave emission by the black hole
having possibly the same order of magnitude. The paper extends our previous result
obtained in the axisymmetric case\cite{ivanoGRG}.
\par Throughout the paper geometric units $G=c=1$ are used.
\section{Derivation of the solutions\label{sectionII}}
\par In obtaining the metric of a stationary non-axisymmetric boosted Kerr black hole
we adopt a simple and elegant apparatus described in Sthephani et al.\cite{kramer}
(sections $29.1$ and $29.5$) to obtain twisting Petrov D vacuum solutions
of Einstein's equations. This procedure follows Kerr in his original derivation of the
Kerr geometry\cite{kerr0}.
The metric is expressed as
\begin{eqnarray}
\label{eqn1}
ds^2= 2 \omega^1 \omega^2 - 2 \omega^3 \omega^4
\end{eqnarray}
where the $1$-forms $\omega^a$ are given by
\begin{eqnarray}
\nonumber
\omega^1 &=&{\bar{ \omega}}^2 = - d \xi/{\bar{\rho}}P,~~~~\omega^3 = du + L d\xi +{\bar{L}} d{\bar{\xi}},\\
\omega^4 &=& dr + W d\xi +{\bar{W}} d{\bar{\xi}}+ H \omega^3,
\label{eqn2}
\end{eqnarray}
in Bondi-Sachs-type coordinates $(u,r,\xi,{\bar{\xi}})$~\cite{bondi1,sachs,sachs1,newman}, where a bar denotes
complex-conjugation.
The metric functions $L$, $W$, $\rho$, $H$ and $P$ are assumed
to be independent of the time coordinate $u$,
namely, $\partial/\partial u$ is a Killing
vector of the geometry. $P$ is a real function. Einstein's vacuum equations result in (cf. \cite{kramer})
\begin{eqnarray}
\rho^{-1}=-(r+i\Sigma),~~~~W=i~ \partial_{\xi} ~\Sigma \;,
\label{eqn3}
\end{eqnarray}
\begin{eqnarray}
H=\lambda/2- \frac{m r}{r^2+\Sigma^2}\;,
\label{eqn4}
\end{eqnarray}
\begin{eqnarray}
\lambda= 2 P^2 ~{\rm Re} ~(\partial_{\xi}\partial_{\bar \xi}~ \ln P)\;,
\label{eqn5}
\end{eqnarray}
\begin{eqnarray}
\lambda \Sigma+ P^2 {\rm Re}~ (\partial_{\xi}\partial_{\bar \xi}~\Sigma)=0\;,
\label{eqn6}
\end{eqnarray}
\begin{eqnarray}
2i \Sigma=P^2 (\partial_{\bar \xi}~L-\partial_{\xi} ~{\bar L})\;,
\label{eqn7}
\end{eqnarray}
where $m$ is a real constant parameter and $\lambda=\pm 1$ is the curvature of the 2-dim surface
$d \xi d{\bar{\xi}}/P^2$. Here $\lambda=1$ is adopted. The $r$-dependence is isolated in $\rho$
and $H$ so that the remaining functions to be determined {\bf --} $P$, $\Sigma$ and $L$ {\bf --} are functions of
$(\xi,{\bar{\xi}})$ only. By a coordinate transformation we have set the origin of the affine parameter $r$
in eqs. (\ref{eqn3}) and (\ref{eqn4}) equal to zero\cite{kramer}.
The remaining field equations to be integrated reduce then to eqs. (\ref{eqn5}), (\ref{eqn6}) and (\ref{eqn7}).
Here we will substitute the variables $(\xi,{\bar \xi})$ by $(\theta,\phi)$ via the stereographic transformation
\begin{eqnarray}
\nonumber
\xi=\cot (\theta/2) e^{i\phi}\;.
\label{eqn8}
\end{eqnarray}
\par The real function $P(\xi,{\bar \xi})$ is integrated from (\ref{eqn5}) by assuming $P$ with the form
\begin{eqnarray}
P=\frac{K(\theta,\phi)}{\sqrt{2}\sin^2 (\theta/2)}\;.
\label{eqn9}
\end{eqnarray}
Eq. (\ref{eqn5}) reduces then to
\begin{eqnarray}
1=K K_{\theta \theta}+K K_{\theta} \cot \theta-K^{2}_{\theta}+K^2+\frac{(K K_{\phi\phi}-K^{2}_{\phi})}{ \sin^2 \theta}\;.~~
\label{eqn10}
\end{eqnarray}
$K(\theta,\phi)=1$ is a solution of (\ref{eqn10}) and corresponds to the original Kerr solution.
As we will see in the following the general $K$-function is the proper and natural tool to introduce the
boost in asymptotically flat gravitational fields, preserving the asymptotic boundary conditions at future
null infinity, even for radiating fields. The $K$-function actually belongs to
the asymptotic orthochronous inhomogeneous Lorentz group that is isomorphic to
conformal transformations of the 2-sphere into itself,
denoted the Bondi-Metzner-Sachs (BMS) group\cite{bondi1,sachs,sachs1,sachs2,penrose}.
\par
A general solution of (\ref{eqn10}) is given by
\begin{eqnarray}
\label{eqn10i}
K(\theta,\phi)= a + b~ {\bf \hat{x}. \bf {n}}\;,~~~~~a^2-b^2=1\;,
\end{eqnarray}
where $\bf {\hat{x}}=(\cos \theta, \sin \theta \cos \phi,\sin \theta \sin \phi)$ is the unit
vector along an arbitrary direction $\bf{x}$ and ${\bf{n}}=(n_1,n_2,n_3)$ is a constant unit vector
satisfying
\begin{eqnarray}
\label{paramN}
n_1^2+n_2^2+n_3^2=1.
\end{eqnarray}
The solution (\ref{eqn10i}) has three independent parameters and defines a general Lorentz boost $K$ contained in the
homogeneous Lorentz transformations of the BMS group. The boost parameter $\gamma$ parametrizes $a$ and $b$
as $(a=\cosh \gamma,b=\sinh \gamma)$, and is associated with the velocity $v=\tanh \gamma$ of the black hole
relative to a Lorentz frame at future null infinity. The case of a non-boosted solution (the Kerr solution) would correspond
to $b=0$.
\par Assuming $\Sigma=\Sigma(\theta,\phi)$  eq. (\ref{eqn6}) in the variables $(\theta,\phi)$ has the form
\begin{eqnarray}
\Sigma_{\theta \theta}+\cot \theta~ \Sigma_{\theta}+\frac{1}{\sin^2 \theta} \Sigma_{\phi\phi}+\frac{2 \Sigma}{K^2(\theta,\phi)}=0\;,
\label{eqn13}
\end{eqnarray}
from which the regular solution is derived,
\begin{eqnarray}
\Sigma(\theta,\phi)= \omega~\frac{b+a({\bf \hat{x}. \bf {n}})}{a+b({\bf \hat{x}. \bf {n}})}\;,
\label{eqn14}
\end{eqnarray}
where $\omega$ is an arbitrary constant to be identified with the rotation
parameter of the solution; the parameters ${\bf {n}}=(n_1, n_2, n_3)$ satisfy (\ref{paramN}).
For $n_2$ and/or $n_3$ non-zero, the black hole solution will be
non-axisymmetric, namely,  $\partial/\partial \phi$ is not a Killing vector of the geometry.
\par Eq. (\ref{eqn7}) can now be integrated using (\ref{eqn14}). Adopting accordingly
\begin{eqnarray}
L(\theta,\phi)=i \mathcal{L}(\theta,\phi) e^{-i \phi}\;,
\label{eqn15}
\end{eqnarray}
where ${\mathcal{L}}(\theta,\phi)$ is real, it results in
\begin{eqnarray}
{\mathcal{L}}_{\theta}- {\mathcal{L}}/\sin \theta + (1-\cos \theta)~\frac{\Sigma(\theta,\phi)}{K^2(\theta,\phi)}=0 \;.
\label{eqn16}
\end{eqnarray}
A general solution of (\ref{eqn16}) is given by
\begin{eqnarray}
\label{eqn17}
{\mathcal{L}}(\theta,\phi)=\Big( \frac{1-\cos\theta}{\sin \theta}\Big) \Big [ C_1- \int \frac{\Sigma(\theta,\phi)}{K^2(\theta,\phi)}~ \sin \theta d\theta \Big ],~~
\end{eqnarray}
where $C_1$ is an arbitrary constant associated with the solution of the homogeneous part of (\ref{eqn16}).
Actually $C_1$ can be an arbitrary function of $\phi$ which however can be re-scaled
to a constant in the final form of the metric. Furthermore in order to avoid an apparent singular
behavior for a zero boost $b^2=0$, as occurring in the axisymmetric Kerr boosted case\cite{ivanoGRG},
the choice $C_1=\omega/2b^2$ is adopted in the remaining of the paper.
For the general boost (\ref{eqn10i}) the integrals in (\ref{eqn17}) are expressed
\begin{eqnarray}
\label{eqn17i}
\int \frac{\Sigma(\theta,\phi)}{K^2(\theta, \phi)} ~\sin \theta~ d \theta= \omega~(I_1+I_2+I_3+I_4),
\end{eqnarray}
where
\begin{eqnarray}
\label{eqn17ii}
\nonumber
&&I_1=  b \int \frac{\sin \theta}{K^3(\theta, \phi)} ~d \theta,~~I_2=  a n_1 \int \frac{\sin \theta \cos \theta}{K^3(\theta, \phi)}~d \theta,\\
&&I_3+I_4= a~ (n_2 \cos \phi+n_3 \sin \phi) \int \frac{\sin^2 \theta}{K^3(\theta, \phi)} ~d \theta.
\end{eqnarray}
These integrals furnish (by the use of a symbolic manipulation package, as Maple) a closed solution in terms of involved combinations
of trigonometric functions, and will not be displayed here for lack of space. In particular, in the axisymmetric case, we obtain consistently
\begin{eqnarray}
\label{eqn17iii}
\int \frac{\Sigma(\theta,\phi)}{K^2(\theta, \phi)} ~\sin \theta~ d \theta= \frac{\omega}{2b^2}~\frac{a^2+2 a b \cos \theta +b^2}{(a+ b \cos \theta)^2}\;.
\end{eqnarray}
\par
The integrands of the integrals in (\ref{eqn17ii}) allows us to define a
Bondi-Sachs {\it 4-momentum aspect} as
\begin{eqnarray}
\label{eqn17iv}
p^{\mu}(\theta,\phi)=\frac{m~ k^{\mu}}{K^3(\theta,\phi)}\;,
\end{eqnarray}
where $k^{\mu}=(-1, \cos \theta,\sin \theta \cos\phi,\sin \theta \sin\phi)$ defines the generators of the BMS
translations in the temporal, and Cartesian directions $x$, $y$ and $z$ of an asymptotic
Lorentz frame at future null infinity, and where $K(\theta,\phi)$ given in (\ref{eqn10i}) is the generator of
Lorentz boosts of the BMS\cite{{bondi1},sachs,sachs1,sachs2}. The integration of (\ref{eqn17iv}) in
the whole sphere yields the total Bondi-Sachs $4$-momentum associated with the the solution (\ref{eqn22}) below,
\begin{eqnarray}
\label{eqn17ivv}
P^{\mu}=\frac{1}{4\pi} \int _{0}^{2 \pi} d\phi \int _{0}^{\pi} p^{\mu}(\theta,\phi) \sin \theta d \theta,
\end{eqnarray}
namely, the Bondi-Sachs mass $M_{BS}$ and the Bondi-Sachs momentum ${\bf P}_{BS}$,
\begin{eqnarray}
\label{eqnBM}
\nonumber
M_{BS}&=&m a=\frac{m} {\sqrt{(1-v^2)}}~,\\
{\bf P}_{BS}&=&m b ~{\bf {n}}=\frac{m v}{{\sqrt{(1-v^2)}}}~~{\bf {n}}.
\end{eqnarray}
The evaluation of (\ref{eqn17ivv}) leading to the result (\ref{eqnBM}) involved a long
and careful computation using the Mathematica package.
\par As we will discuss below the mass and momentum aspects (\ref{eqn17iv}) are physical quantities that
contribute to the angular momentum of the solution.
\par
From (\ref{eqn15}) we obtain
\begin{eqnarray}
\label{eqn18}
L d{\xi}+ {\bar L}{\bar \xi}=-2~ {\mathcal{L}}({\theta, \phi}) ~\cot \theta/2~d \phi \;,
\end{eqnarray}
where $ {\mathcal{L}}(\theta,\phi)$ is given by (\ref{eqn17}).
\par
Analogously from (\ref{eqn1}), $W=i \partial_{\xi}~ \Sigma$, it results
\begin{eqnarray}
\label{eqn19}
W(\theta,\phi)= e^{-i\phi} \Big[-i \Sigma_{\theta}~ \sin^2 \theta/2+\frac{\Sigma_{\phi}}{2\cot \theta/2} \Big]\;,
\end{eqnarray}
and we obtain
\begin{eqnarray}
\label{eqn20}
\nonumber
&&W d{\xi}+ {\bar W} d{\bar \xi}= \Big[ \Sigma_{\theta}~ \sin \theta~d\phi -\frac{\Sigma_{\phi}}{\sin \theta}~d \theta \Big]\\
\nonumber
&&=\omega\Big( \frac{-n_1 \sin^2 \theta+ (n_2 \cos \phi+n_3 \sin \phi)\sin \theta \cos \theta}{K^2(\theta,\phi)}\Big)~ d\phi\\
&&+~\omega\frac{(n_2 \sin \phi -n_3 \cos \phi)}{K^2(\theta,\phi)} ~d \theta
\;.
\end{eqnarray}
In order to complete the metric 1-forms (\ref{eqn2}) we have
\begin{eqnarray}
\label{eqn21}
H= \frac{1}{2}- \frac{m r}{r^2+\Sigma^2(\theta,\phi)}\; ,~~~\rho^{-1}= -(r+i \Sigma(\theta,\phi))\;.
\end{eqnarray}
The metric (\ref{eqn1}) finally results
\begin{widetext}
\begin{eqnarray}
\label{eqn22}
\nonumber
ds^2= \frac{r^2+\Sigma^2(\theta,\phi)}{K^2(\theta,\phi)}(d\theta^2+\sin^2 \theta ~d \phi^2)
-2(du-2 {\mathcal{L}(\theta,\phi)} \cot \theta/2 ~ d\phi)\times\\
\nonumber
\Big[ dr + \omega\frac{-n_1 \sin^2 \theta+(n_2 \cos \phi+ n_3 \sin \phi)\sin \theta \cos \theta}{K^2(\theta,\phi)} ~d\phi
\nonumber
+\omega\frac{n_2 \sin \phi- n_3 \cos \phi}{K^2(\theta,\phi)}~d\theta \Big]\\
-(du-2 {\mathcal{L}(\theta,\phi)} \cot \theta/2 ~ d\phi)^2 \times \frac{r^2-2mr+\Sigma^2(\theta,\phi)}{r^2+\Sigma^2(\theta,\phi)}\;.
\end{eqnarray}
\end{widetext}
where $K(\theta,\phi)$ and $\Sigma(\theta,\phi)$ are given in (\ref{eqn10i}) and (\ref{eqn14}), respectively,
and ${\mathcal{L}(\theta,\phi)}$ in ({\ref{eqn17}}).
The metric describes a boosted Kerr black hole along an arbitrary direction relative an
an asymptotic Lorentz frame at future null infinity. The direction of the boost
is defined by the Euler parameters $(n_1, n_2, n_3)$, cf. (\ref{eqn10i}), of the Lorentz
boosts of the BMS group\cite{bondi1}.
\par
For $n_2=0=n_3$ and $b=0$ the metric (\ref{eqn22}) is the original Kerr
metric in retarded Bondi-Sachs-type coordinates\footnote {We note that these coordinates
correspond actually to the standard Kerr-Schild or Eddington-Finkelstein coordinates
used largely in the literature of Kerr spacetimes.}. For $\omega=0$ it represents a boosted Schwarzschild
black hole along the direction determined by $(n_1, n_2, n_3)$.
\begin{figure*}
\begin{center}
\hspace{0.5cm}
\vspace{0.0cm}
\includegraphics[width=7.0cm,height=7.6cm]{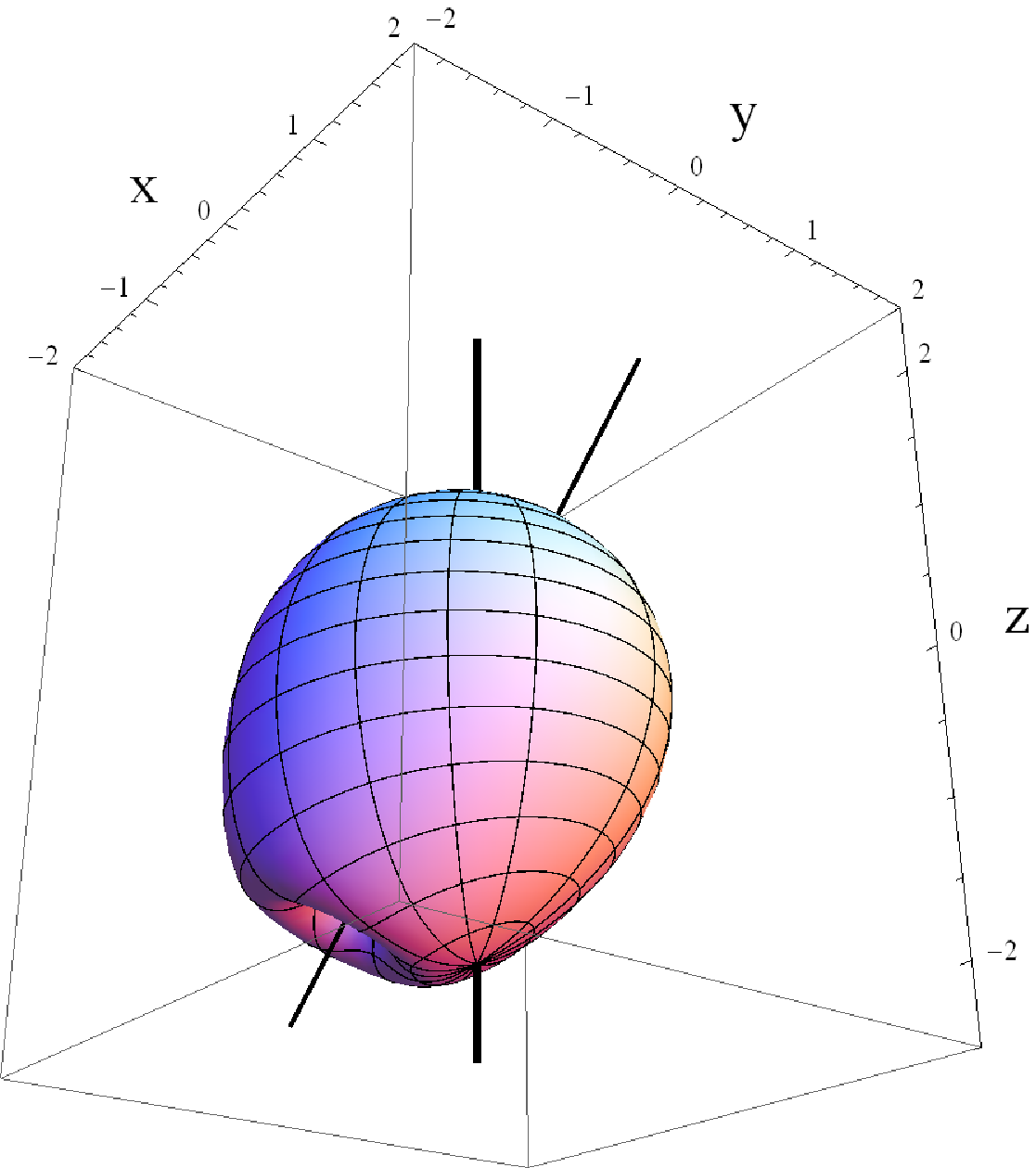}\includegraphics[width=7.6cm,height=7.6cm]{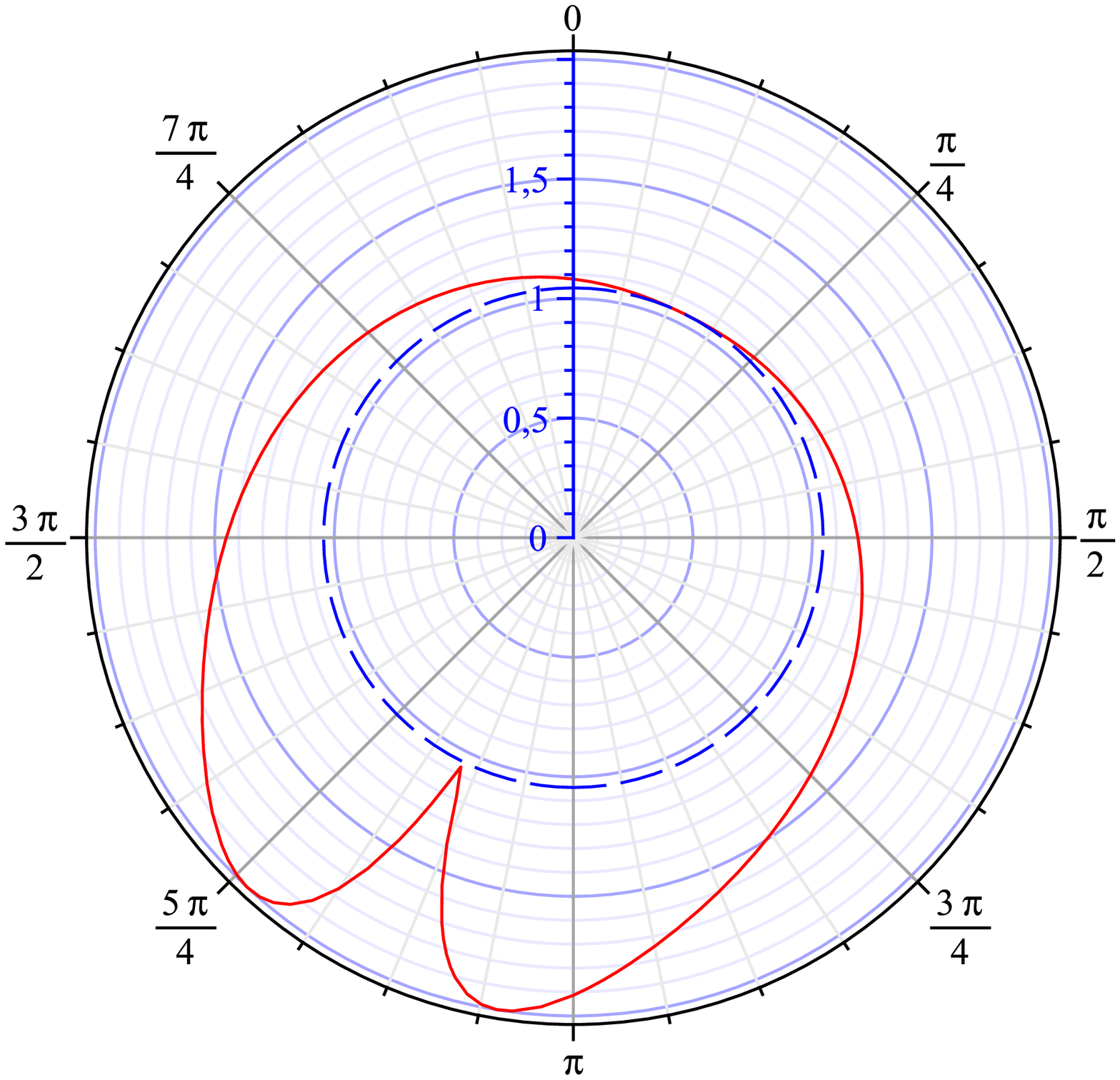}
\caption{Plots of the ergosphere for a boosted Kerr black hole, with $m=1$, $\omega=0.999$,
and the boost direction taken as $n_1=0.9$, $n_2=0.3$, $n_3 \simeq 0.316227766$. The boost parameters $b$
and $a$ correspond to the velocity of the black hole $v/c\simeq 0.956532$. (left) A 3-dim
view of the ergosphere static limit where the thin black axis corresponds to the direction $(n_1,n_2,n_3)$ of the boost while
the thick black axis is the axis of rotation (the $z$-axis) of the black hole. (right) A section of the ergosphere static limit
by the plane $\phi=\pi/3.931$ (red continuous line) shown in the plane $\theta$. The boost axis shown in the left figure
is defined by the two points $\theta \simeq 25.84^{\circ}$ and $\theta \simeq 25.84^{\circ} \pm \pi$ where the ergosphere
contacts the 2-dim event horizon $r_{+}\simeq1.044710$ (dashed blue line).
}
\label{ergoB}
\end{center}
\end{figure*}
\begin{figure*}
\begin{center}
\hspace{0.0cm}
\vspace{0.0cm}
\includegraphics[width=5.6cm,height=5.6cm]{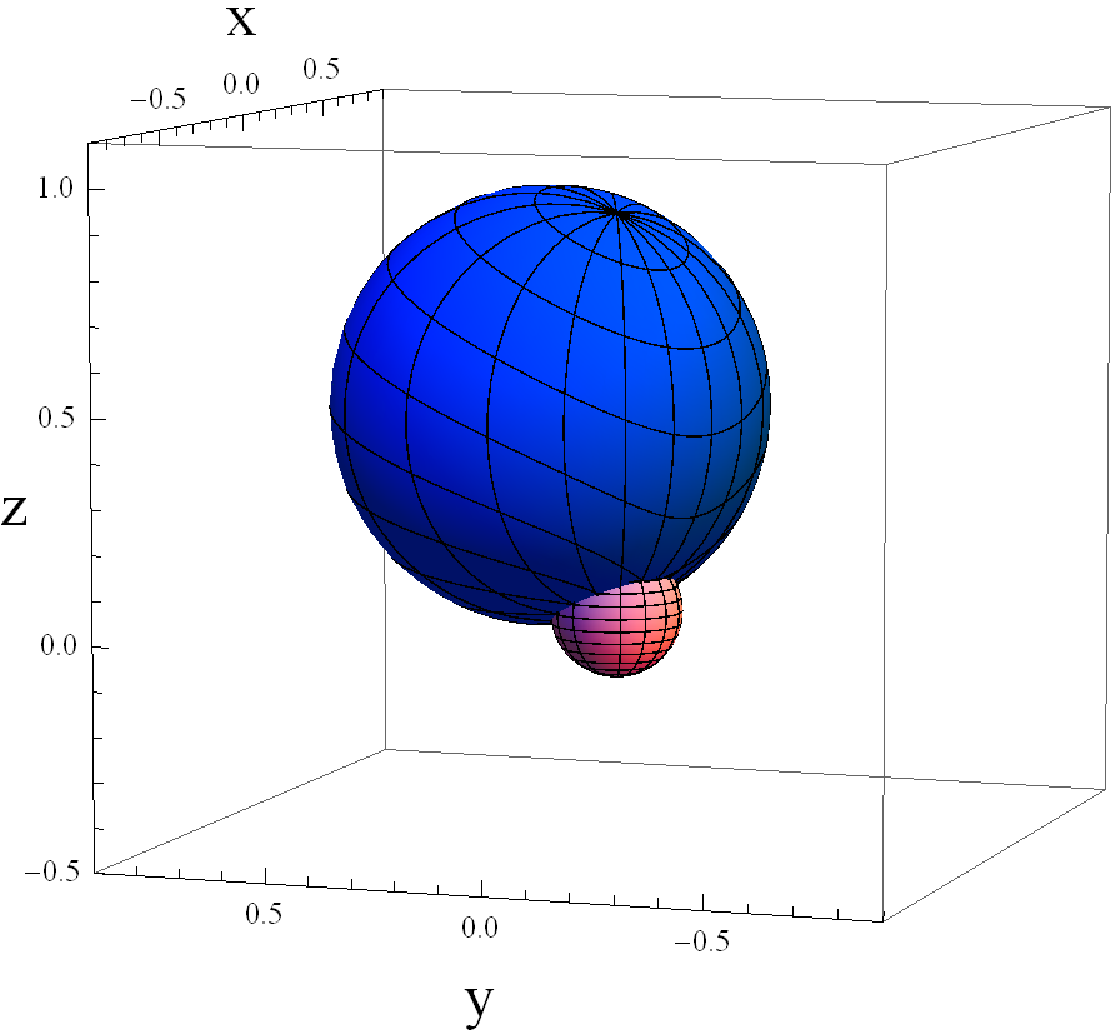}~~~~\includegraphics[width=5.6cm,height=5.6cm]{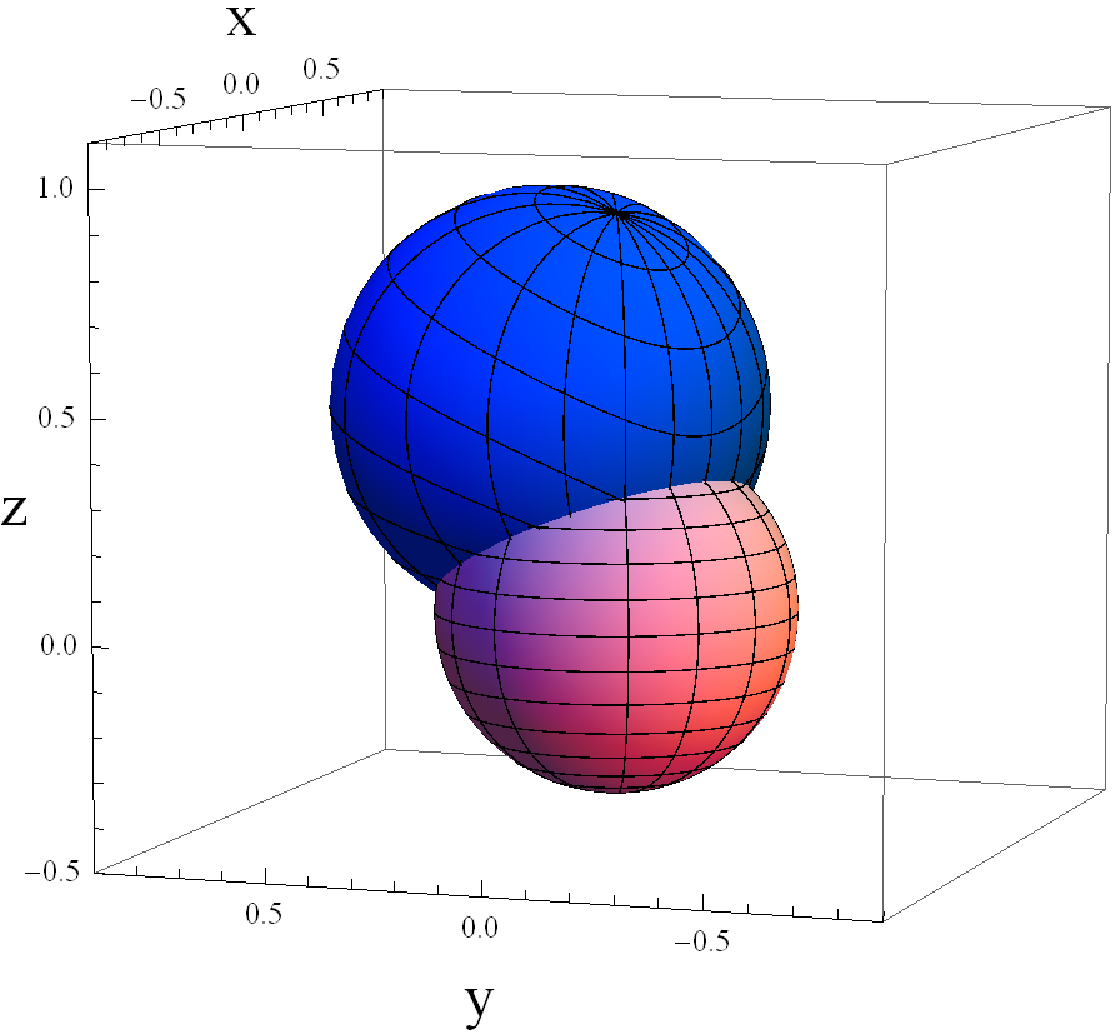}
\includegraphics[width=5.0cm,height=6.6cm]{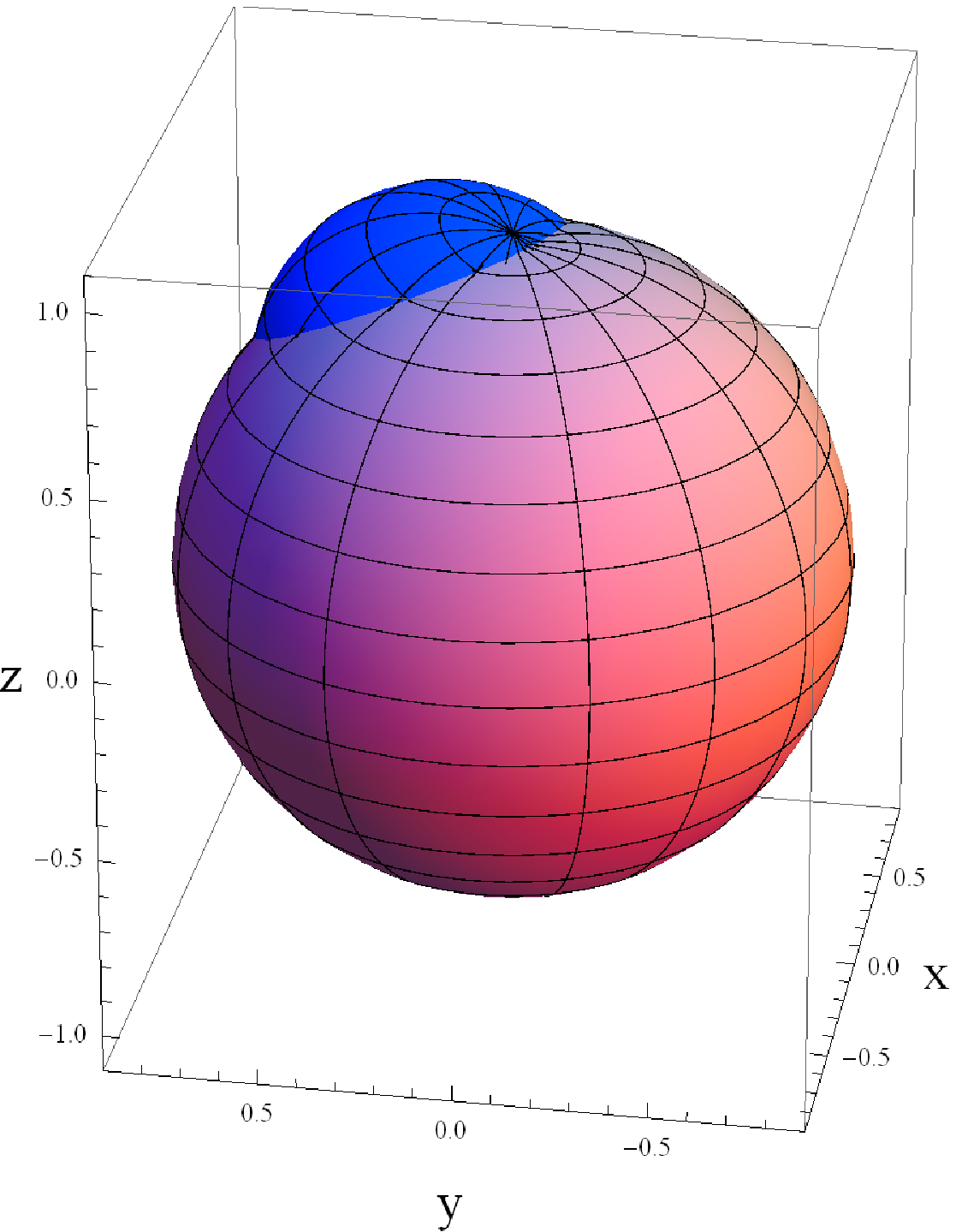}
\caption{Plots of the intersections of the surface $\mathcal{H}(\theta,\phi)=(n_1 \cos \theta+n_2 \sin\theta \cos \phi + n_3 \sin\theta \sin \phi )$
(blue) with the sphere $\mathcal{S}(\theta,\phi)$ about the origin with radius $-(b/a)$ (red). The intersection defines a closed
line of singularities; for increasing values of $|b|$ these closed curves increase and then decrease as the radius of the spherical
surface (red) about the origin increases. The figures correspond to
the three values of $b=-0.15$, $b=-0.45$ and $b=-1.8$ (from left to right), with fixed
parameters $(n_1=0.9,n_2=0.2,n_3=\sqrt{1-n_1^2-n_2^2})$. For $b=0$ and $b \rightarrow \pm \infty$ the circles
reduce to a point. The closed singularity curves are contained in planes orthogonal
to the direction of the boost $(n_1,n_2,n_3)$.
}
\label{singularity}
\end{center}
\end{figure*}
\begin{figure*}
\begin{center}
\hspace{0.0cm}
\vspace{0.0cm}
\includegraphics[width=5.8cm,height=5.8cm]{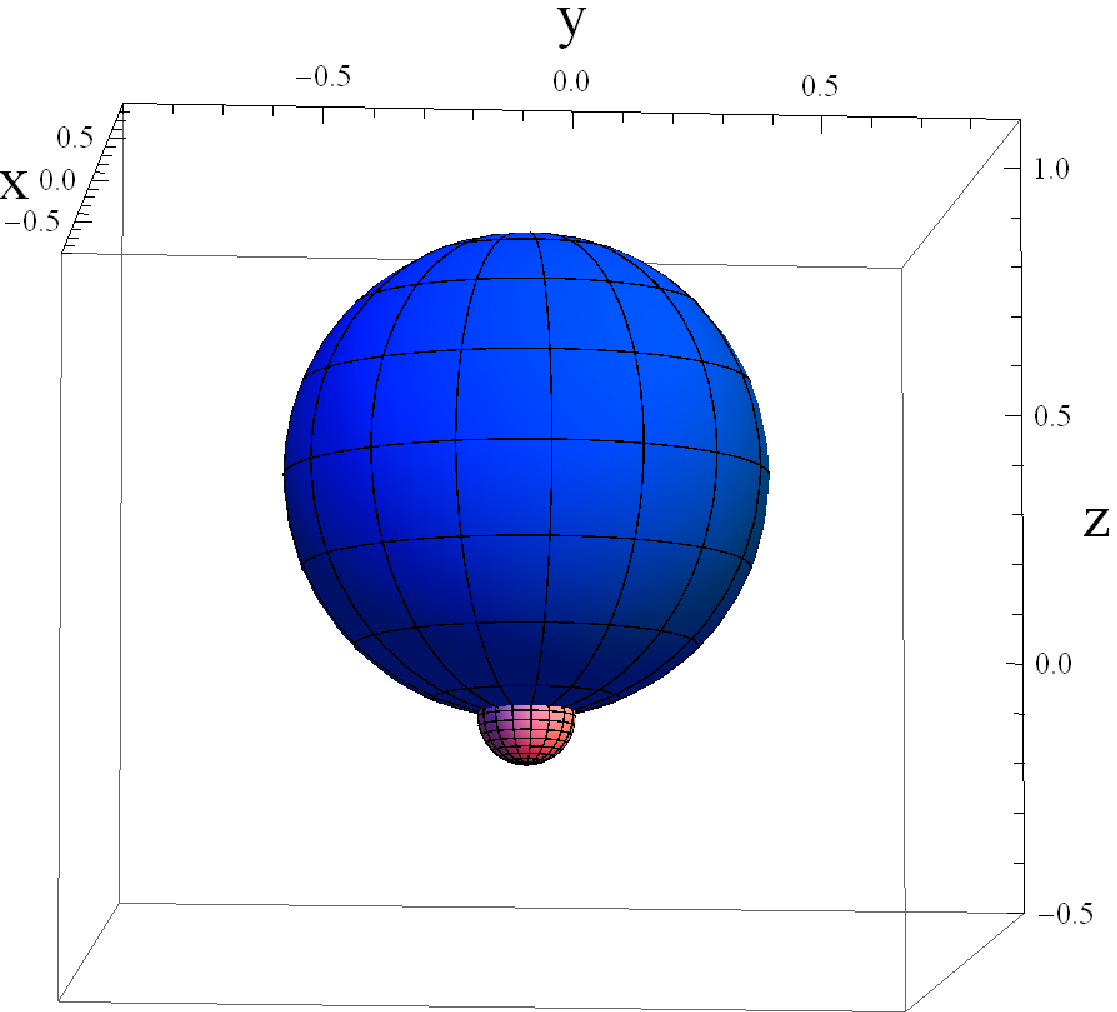}\includegraphics[width=5.8cm,height=5.8cm]{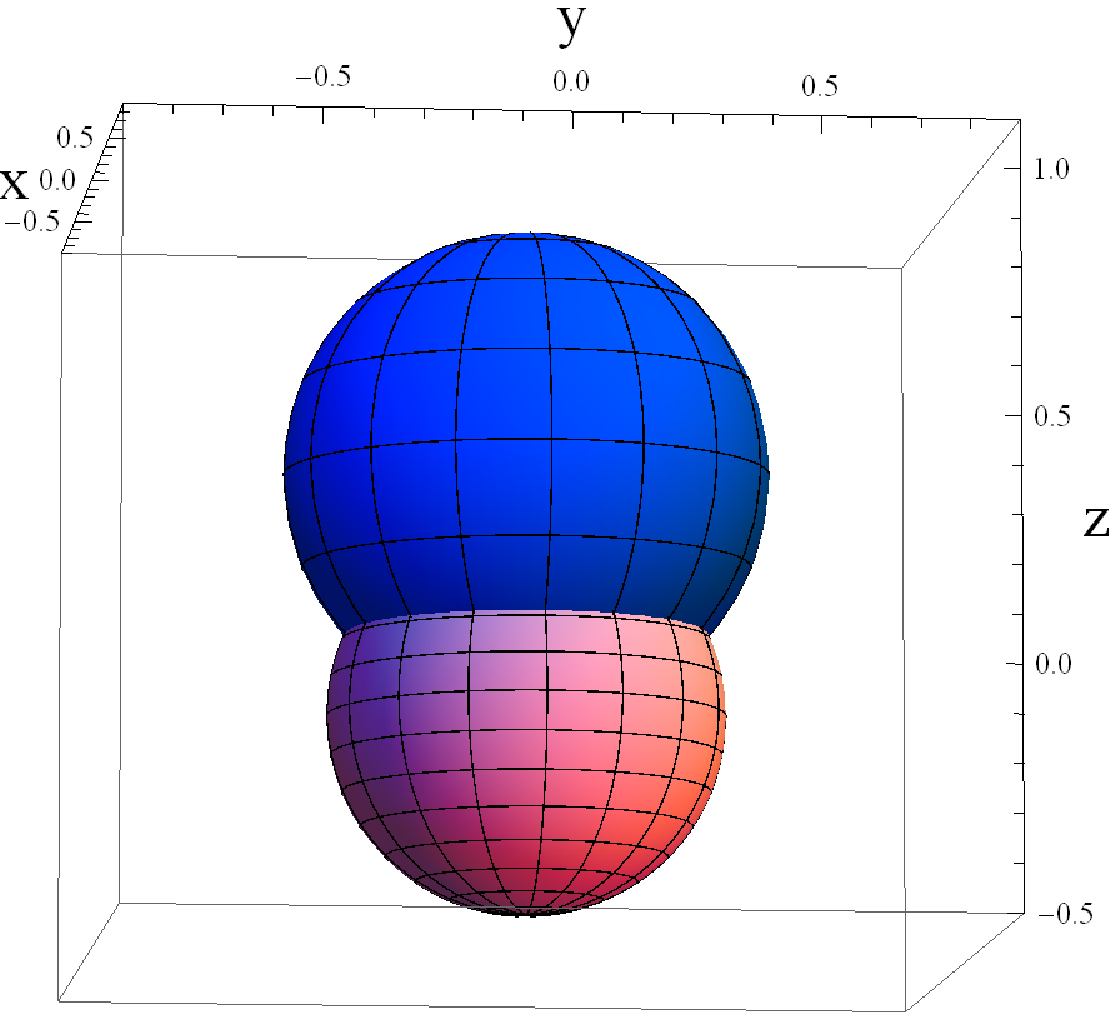}
\includegraphics[width=5.8cm,height=5.8cm]{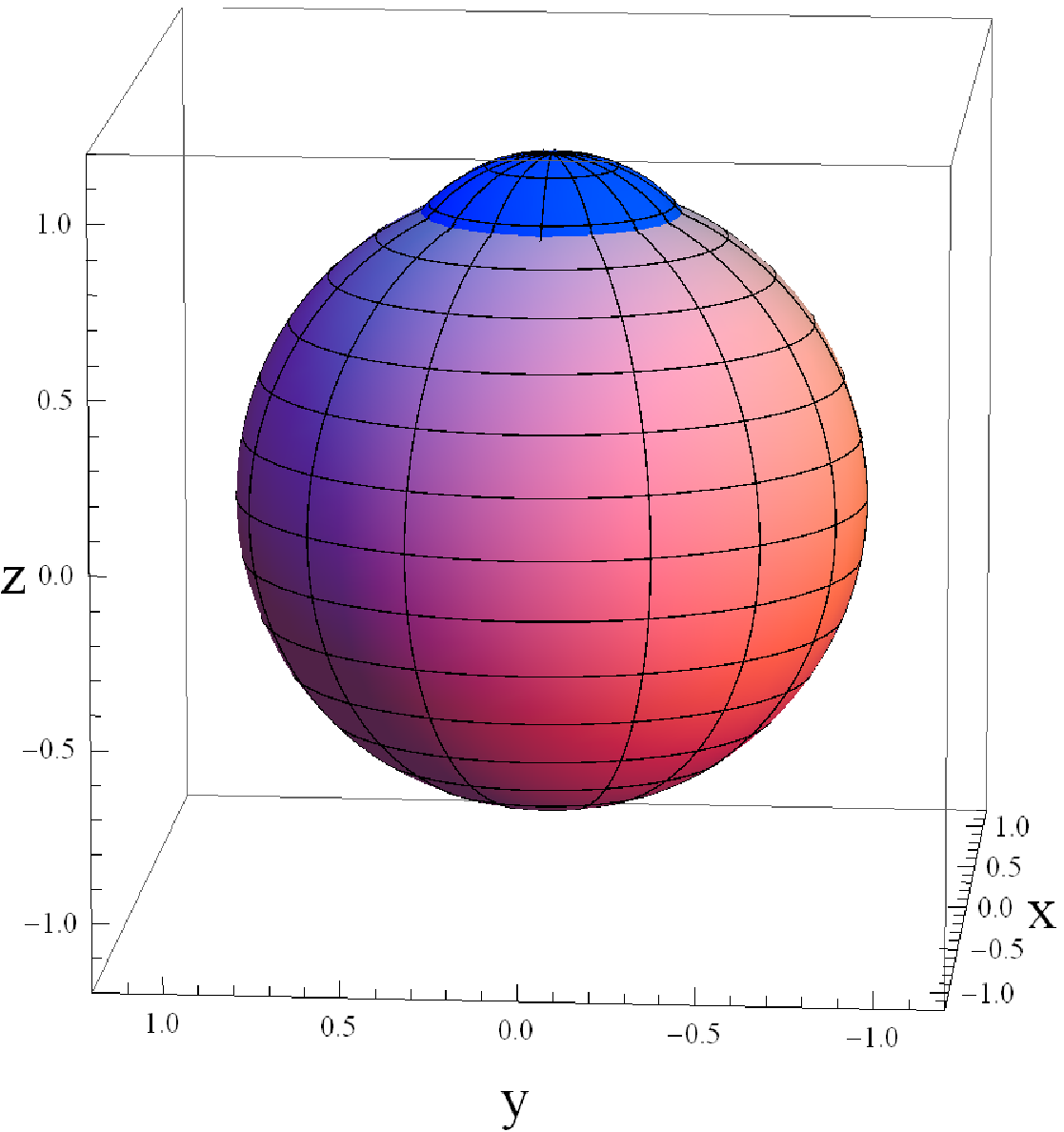}
\caption{Plot of the intersection of the surface $\mathcal{H}(\theta,\phi)$ (blue) and the sphere $\mathcal{S}(\theta,\phi)$ (red)
for the axial boosted case, with $n_1=1$, $n_2=0=n_3$ and $b=-0.1$, $b=-0.45$ and $b=-1.2$ (from left to right).
The intersections are circles of singularity on the $z={\rm const.}$ (or $\theta={\rm const.}$) planes,
as should be expected in the axial configuration. For $b=0$ and $|b| \rightarrow  \infty$ the circles
reduce to a point respectively at the equatorial plane $\theta=\pi/2$ and at the poles $\theta=0,~\pi$.}
\label{singularity_Ax}
\end{center}
\end{figure*}
\par By isolating the mass dependent term in the above geometry we obtain
\begin{eqnarray}
\label{eqn24}
ds^2=ds^2_M+ \frac{2mr}{r^2+\Sigma^2(\theta,\phi)}~(l_{\alpha} dx^{\alpha})^2\;,
\end{eqnarray}
where $l_{\alpha}=(1,0,0,-2 \mathcal{L}(\theta,\phi) \cot \theta/2 )$ is a null vector
with respect to both metrics $ds^2$ and $ds^2_M$, namely, $l_{\alpha}l^{\alpha}=0$.
In verifying these results we used
\begin{eqnarray}
\label{eqn26}
\nonumber
g^{uu}=\frac{4 K^2(\theta,\phi)~\mathcal{L}^2(\theta,\phi)}{(\cos \theta-1)^2 ~(r^2+\Sigma^2(\theta,\phi))}\;,\\
\nonumber
g^{u\phi}=-\frac{2 K^2(\theta,\phi)~\mathcal{L}(\theta,\phi)}{(\cos \theta-1)~\sin \theta (r^2+\Sigma^2(\theta,\phi))}\;,\\
\nonumber
g^{\phi\phi}=\frac{4 K^2(\theta,\phi)}{\sin^2 \theta~ (r^2+\Sigma^2(\theta,\phi))}\;.
\end{eqnarray}
\par The metric $ds^2_M$ does not involve the mass and has the associated Riemann tensor equal to zero,
as can be tested carefully, being the metric $g_{(M){\alpha \beta}}$ of a flat space, so that (\ref{eqn22})
assumes the Kerr-Schild form
\begin{eqnarray}
\nonumber
g_{\alpha \beta}=g_{(M){\alpha \beta}}+\frac{2mr}{r^2+\Sigma^2(\theta,\phi)}~l_{\alpha}l_{\beta}.
\end{eqnarray}
As we are in the realm of Kerr metrics it is straightforward to see that in the axisymmetric case,
namely, when $n_2=0=n_3$, the metric (\ref{eqn22}) reduces either to the axial Kerr boosted metric
or to the Kerr metric, whether $b \neq 0$ or $b=0$ respectively, the metrics having the Killing vectors
$\partial/\partial u$ and $\partial/\partial \phi$.
\par
Expanding (\ref{eqn22}) in the axisymmetric case for large $r$ and in the limit of slow rotation parameter $\omega \ll m$, we obtain
\begin{eqnarray}
\label{eqn22S}
\nonumber
ds^2 \simeq &-& (1-\frac{2m}{r}) du^2 - 2 du dr +\frac{r^2(d\theta^2+\sin^2 \theta ~d \phi^2)}{K^2(\theta)},\\
&+& \frac {4m \omega}{r K^2(\theta)}~ \sin^2 \theta  ~ du d\phi,
\end{eqnarray}
where $K(\theta)=(a+b\cos\theta)$. This linearized version represents a boosted mass monopole plus the Lense-Thirring
rotating term with angular momentum $m \omega$\cite{lense,janis,poisson}.
Therefore in the axisymmetric case the metric (\ref{eqn22}) can then be interpreted as a
boosted Kerr black hole rotating about the
$z$-axis (the axis defining the angle $\phi$) with angular momentum $m \omega$. The boost
is along the axis of rotation, relative to an asymptotic Lorentz frame at future null infinity.
\par
Now for the general boosted case an analogous expansion results in
\begin{eqnarray}
\label{eqn22SS}
\nonumber
ds^2 \simeq &-& (1-\frac{2m}{r}) du^2 - 2 du dr + \frac{r^2(d\theta^2+\sin^2 \theta ~d \phi^2)}{K^2(\theta,\phi)}\\
&-& \frac {4m}{r}~ {\mathcal{L}(\theta,\phi)} \cot \theta/2 ~ du d\phi.
\end{eqnarray}
By comparing (\ref{eqn22S}) and (\ref{eqn22SS}) we observe that a difference appears in the
Lense-Thirring rotation term.
This actually results from the fact that -- besides the mass aspect -- the rotation
term of (\ref{eqn22S}) contains just the
{\it 3-momentum aspect} component $p^{1}$ of the geometry, while in (\ref{eqn22SS}) the complete
{\it 3-momentum aspect} $(p^{1},p^{2},p^{3})$ is present.
In this sense we have in (\ref{eqn22SS}) a genuine natural extension of the Lense-Thirring rotation term.
The vanishing of the components $p^2$ and $p^3$ of the {\it 3-momentum aspects} (\ref{eqn17iv}),
by taking $n_2=0=n_3$, restores the axisymmetry of the boosted black hole configuration.
We remark that the presence of a momentum aspect that adds to the mass aspect in (\ref{eqn17i}) is mandatory
since we cannot make all the Euler parameters zero due to the relation $n_1^{2}+n_2^{2}+n_3^{2}=1$.
In all cases the angular momentum is proportional to $m \omega$ about the axis defining the coordinate $\phi$.
\par We note that the $1/K^2$-factor multiplying the $2$-sphere line
element in (\ref{eqn22S}) and (\ref{eqn22SS}) is actually a conformal transformation of the unit $2$-sphere into itself
which is isomorfic to a Lorentz boost of the BMS group\cite{sachs2,penrose}.
The $z$-axis about which the black hole rotates does not coincide with the boost axis except in
the axial case $n_2=0=n_3$\cite{ivanoGRG}.
A further detailed examination of the rotation term $du d\phi$ of the Kerr boosted geometry (\ref{eqn22})
will be given in following Section, where the angular momentum of the event horizon $r=r_{+}$ is analyzed.
\par
Finally for illustration and comparison with the above results we present the
slow rotation limit of a non-boosted nonaxisymmetric Kerr black hole (cf. (\ref{eqn22})) that reads
\begin{eqnarray}
\label{eqn22SSS}
\nonumber
&&ds^2 \simeq - (1-\frac{2m}{r}) du^2 - 2 du dr + r^2(d\theta^2+\sin^2 \theta ~d \phi^2)\\
\nonumber
&+& \frac {4m \omega}{r}~\Big(n_1 \sin^2 \theta + (n_2 \cos \phi+n_3 \sin \phi)\times\\
&&(\theta-\sin\theta \cos\theta)  \Big)~ du d\phi.
\end{eqnarray}
The angular momentum at the equator in this case results $\Omega(\phi)=m \omega(n_1+\pi~(n_2 \cos \phi+n_3 \sin \phi)/2)$.

\section{Properties of the solution: the ergosphere and horizons\label{sectionIII}}

A direct examination of (\ref{eqn22}) shows that $\partial/\partial u$ is a Killing vector of
the geometry and defines its stationary character. The general boosted Kerr geometry also presents an
ergosphere, defined by the limit surface for static observers, namely, the locus where the Killing vector
$\partial/\partial u$ becomes null\cite{poisson}, and by the event horizon to be discussed below.
In the coordinate system of (\ref{eqn22}) the equation of the limit surface $g_{uu}=0$ results in
\begin{eqnarray}
\label{eqn27}
r^2 -2mr+\Sigma^2(\theta,\phi)=0\;,
\end{eqnarray}
or
\begin{eqnarray}
\label{eqn28}
r_{stat}(\theta, \phi)= m +\sqrt{m^2-\Sigma^2(\theta,\phi)}  \;.
\end{eqnarray}
The horizons are the surfaces defined by
\begin{eqnarray}
\label{eqn29}
g^{rr}=\frac{1}{{r^2+\Sigma^2(\theta,\phi)}}~\Big[ r^2-2 m r + \omega^2 \frac{\Delta(\theta,\phi)}{K^2(\theta,\phi)}\Big ]=0,~~
\end{eqnarray}
where
\begin{eqnarray}
\label{eqn30}
\nonumber
\Delta(\theta,\phi)=(b + a~ {\bf \hat{x}. \bf {n}})^2 +\Delta_1(\theta,\phi),
\end{eqnarray}
and
\begin{widetext}
\begin{eqnarray}
\nonumber
\Delta_1(\theta,\phi)&=&n_1^2 \sin^2 \theta + n_2^2 (\cos^2 \phi \cos^2 \theta+ \sin^2 \phi) + n_3^2 (\cos^2 \phi \sin^2 \theta +\cos^2 \theta)\\
&-&2 n_1 n_2 (\sin \theta \cos \theta \cos \phi) -2 n_1 n_3 (\sin \theta \cos \theta \sin \phi)-2 n_2 n_3 (\sin^2 \theta \sin \phi \cos \phi )\;.
\label{eqn30i}
\end{eqnarray}
\end{widetext}
Since $(n_1^2+n_2^2+n_2^2)=1$ and $(a^2-b^2)=1$ for a general boost, cf.  (\ref{eqn10i}), it follows after some algebra
that $\Delta(\theta,\phi)/K^2(\theta,\phi)=1$, so that (\ref{eqn29}) reduces to the simple form
\begin{eqnarray}
\label{eqn31}
\nonumber
r^2 -2 m r + \omega^2=0,
\end{eqnarray}
with roots
\begin{eqnarray}
\label{eqn31}
r_{\pm}=m \pm \sqrt{m^2-\omega^2} \;.
\end{eqnarray}
We can see that the event horizon $r_{+}$ and the Cauchy horizon $r_{-}$ do not alter by the effect
of the boost.
\par
The region between the surfaces $r_{stat}$ and $r_{+}$ is the ergosphere where the Penrose
process\cite{penrose1,mtw} takes place. The ergosphere is deformed
along the direction ${\bf{n}}=(n_1,n_2,n_3)$ of the boost, as illustrated in Fig.\ref{ergoB} (left)
for a Kerr black hole with mass $m=1$ and rotation parameter $\omega=0.999$,
in geometrical units. The boost parameter adopted to construct the figure was $b=3.28$, corresponding to
the boost velocity of the black hole $v/c=\tanh \gamma \simeq 0.956532$,
relative to a Lorentz frame at null infinity.
The ergosphere static limit surface contacts the event horizon $r_{+}$ in just two points. In the configuration
of the Fig.\ref{ergoB} (right) these points correspond to $\phi=\pi/3.931$, with $\theta \simeq 25.84^{\circ}$
and $\theta \simeq 25.84^{\circ} \pm \pi$.
\par
Both horizons $r_{+}$ and $r_{-}$ constitute $3$-dim manifolds with the topology of $S^3$,
although the associated geometry is not spherical. This property -- already known for the case of
the original Kerr black hole spacetime\cite{mattvisser,smarr} -- holds also for the boosted extensions of the Kerr manifold.
For the sake of simplicity and space we consider the case of the event horizon $r=r_{+}$ only, and
the restriction to the section $\theta=\pi/2$ of the geometry (\ref{eqn22}). The case of the Cauchy horizon $r_{-}$
has a similar analysis and results. We obtain
\begin{widetext}
\begin{eqnarray}
\nonumber
ds^2|&=& \Omega(\phi) ~du d\phi-\Big(\frac{\Sigma^2(\pi/2,\phi)-\omega^2}{r_{+}^{2}+\Sigma^2(\pi/2,\phi)} \Big)du^2\\
&&+\Big(\frac{r_{+}^{2}+\Sigma^2(\pi/2,\phi)-4\mathcal{L}(\pi/2,\phi)\omega n_1}{K^2(\pi/2,\phi)}- 4\mathcal{L}(\pi/2,\phi) [\frac{\Sigma^2(\pi/2,\phi)-\omega^2}{r_{+}^{2}+\Sigma^2(\pi/2,\phi)}]\Big) d\phi^2
\;,
\label{eqn22vv}
\end{eqnarray}
\end{widetext}
where
\begin{eqnarray}
\label{eqn22v4}
\nonumber
&&\Omega(\phi)=2 \omega\Big(\frac{ n_1}{K^2(\pi/2,\phi)}+\frac{2\mathcal{L}(\pi/2,\phi)(\Sigma^2(\pi/2,\phi)-\omega^2)}{\omega~(r_{+}^{2}+\Sigma^2(\pi/2,\phi))} \Big),\\
\nonumber
&&\Sigma^{2}(\pi/2,\phi)-\omega^2= \omega^2~\frac{(n_2 \cos \phi+n_3 \sin\phi)^2-1}{K^2(\pi/2,\phi)},\\
&&K(\pi/2,\phi)=a+b~(n_2 \cos \phi+n_3 \sin\phi)\;.
\end{eqnarray}
The term $\Omega(\phi)$ of the restricted geometry (\ref{eqn22vv})
corresponds to a rotation about the $z$-axis. The
associated angular momentum is not conserved since
$\partial/\partial \phi$ is
not a Killing vector of the black hole geometry. Furthermore from Fig. \ref{ergoB} we also can see that
the angle between the boost axis, defined by the Euler parameters $(n_1, n_2, n_3)$, and the rotation axis $z$
is $\theta_0=\arccos (n_1)$.
\par If we consider the axisymmetric limit of (\ref{eqn22}), the rotational term of
the event horizon $r=r_{+}$ at the equatorial plane $\theta=\pi/2$ reduces to
\begin{eqnarray}
\label{eqnHorizon2}
\Omega=\omega~\frac{4m r_{+}}{(a^2 r^{2}_{+}+\omega^2 b^2)}.
\end{eqnarray}
Finally the geometry of the event horizon at the section $du=0$ results in
\begin{eqnarray}
\label{eqnHorizon3}
\nonumber
ds^2|=\frac{r_{+}^2+\Sigma^2(\theta)}{(a+b \cos \theta)^2}~(d\theta^2+\sin^2 \theta~ d\phi^2)\\
+\frac{\omega^2 \sin^4 \theta}{(a+b \cos \theta)^4}~\Big( \frac{r_{+}^2+\Sigma^2(\theta)+ 2m r_{+}}{r_{+}^2+\Sigma^2(\theta)}\Big) d\phi^2,
\end{eqnarray}
which is topologically, but not geometrically, a $2$-sphere. For the general boosted Kerr black hole this
property can be checked numerically.

\section{The singularity\label{sectionIV}}

By a careful examination of the curvature invariants of (\ref{eqn22}) we
can see that the metric and the curvature are truly singular at
\begin{eqnarray}
\label{eqn32}
r^2+ \Sigma^2(\theta,\phi)=0  \;,
\end{eqnarray}
namely, at
\begin{eqnarray}
\label{eqn33}
r=0,~~ \Sigma(\theta,\phi)=0  \;.
\end{eqnarray}
The singularity is then contained in the $2$-dim surface defined by
\begin{eqnarray}
\label{eqn34}
(n_1 \cos \theta+ n_2 \sin \theta \cos \phi + n_3 \sin \theta \sin \phi)= -b/a  \;,
\end{eqnarray}
at $r=0$. Specifically the singular points correspond to closed curves which are the
intersection of the $2$-dim surface $\mathcal{H}(\theta,\phi)$ with the
$2$-sphere $\mathcal{S}(\theta,\phi)=b/a$ with center at the origin,
\begin{eqnarray}
\label{eqnH1}
\mathcal{H}(\theta,\phi)=(n_1 \cos \theta+ n_2 \sin \theta \cos \phi + n_3 \sin \theta \sin \phi)\;,
\end{eqnarray}
\begin{eqnarray}
\label{eqnH2}
\mathcal{S}(\theta,\phi)=-b/a, ~~\rm{for~ all}~(\theta,\phi)\;.
\end{eqnarray}
For increasing values of $|b|$ -- as the radius of the (red) sphere $\mathcal{S}(\theta,\phi)$
about the origin increases -- the radius of the singularity lines initially increases and then decreases.
In the limits $b=0$ and $b \rightarrow \infty$ (that is, when $b/a=v/c\rightarrow 1$)
the closed curves reduce to a point. This is illustrated in Figs. \ref{singularity} for of $b=-0.15,~-0.45~\rm{and}~ -1.8$.
\par For the axisymmetric boosted case ($n_1=1$ and $n_2=0=n_3$) the closed curves
corresponding to the singularity of the black hole are circles
on the planes $z=\rm{const}$ (namely $\theta_S=\rm{arccos} (-b/a)$~), with $0<|b|< \infty$, as
illustrated in Fig. \ref{singularity_Ax} for $b=-0.1$, $b=-0.45$ and $b=-1.2$.
Analogous to the general boosted case, for $b \rightarrow \pm \infty$ the singularity circle reduces
to a point at the north/south poles $(\theta=0,~\pi)$;
~for $b=0$ (the Kerr black hole) the circle reduces
to a point on the equatorial plane.
\par The point singularities occurring at the equatorial plane and at the north/south poles
(corresponding respectively to $b=0$ and  $b \rightarrow \pm \infty$) can be developed by using
the flat metric $ds^2_{M}$ of the Kerr-Schild form (\ref{eqn24}). A straightforward calculation
yields that in the axial case ${\mathcal{L}(\theta,\phi)}_{S}=-\omega ~ (a+b)/2$
so that the background geometry $ds_M^2$ in (\ref{eqn24}) assumes the form
\begin{eqnarray}
\label{SingI}
\nonumber
{ds_{M}^2}{\Big{|}}_{S} = -du^2 + \omega^2 ~d \phi^2\;,
\end{eqnarray}
independent of $b$, so that in terms of the Minkowski metric $ds^2_M$
the curvature singularity at $\theta=\pi/2$ and $\theta=0,~\pi$ has the topology of a ring\cite{mattvisser}.
This pattern is to be maintained for the general boosted case except that the closed singularity curves are
obviously no longer on the equatorial plane, being actually contained in planes
orthogonal to the direction of the boost determined by $(n_1,n_2,n_3)$.

\section{Discussions and Conclusions}

In this paper we have derived a solution of Einstein's vacuum equations (\ref{eqn22})
corresponding to a general boosted Kerr black hole that
describes the most general configuration of a remnant astrophysical black hole present in nature.
\par
Astrophysical processes in which black holes are formed were the object of recent detections by the
LIGO Scientific Collaboration and the Virgo Collaboration\cite{ligo1}, of the gravitational
waves emitted by a binary black hole merger\cite{gw1,gw2,gw3,gw4} with mass ratios
in the range $0.53-0.83$. The unequal masses of the initial black holes in the observed binaries
result that the gravitational waves emitted have a non-zero gravitational wave momentum flux,
indicating that the remnant black hole must be a Kerr black hole boosted along the direction of the
late time momentum flux, with respect to the asymptotic Lorentz frame at null infinity where such
emissions have been detected.
The remnant black hole solution has five independent parameters, namely, the mass $m$,
the rotation parameter $\omega$, the boost velocity $v=\tanh \gamma$ and the direction of the boost
determined by $(n_1,n_2,n_3)$ satisfying $(n_1)^2+(n_2)^2+(n_3)^2=1$.
These parameters are necessary to the description of a remnant black hole in nature.
In the integration of the solution the general $K(\theta,\phi)$-function appears as the appropriate
and natural tool to introduce the boost in asymptotically flat gravitational fields,
preserving the asymptotic boundary conditions at future null infinity.
The additional parameters connected to the boost do not change the Kerr black hole structure, namely,
the event and the Cauchy horizons, both having the topology of a $3$-sphere.
The paper extends our previous results
obtained in the axisymmetric case\cite{ivanoGRG}.
\par The issue of the rotation of the black hole is examined firstly in the case of large $r$
and the slow rotation limit. We obtain that this linearized version represents
a boosted mass monopole plus the Lense-Thirring rotating term with angular momentum proportional
to $m \omega$ about the axis defining the angle $\phi$. By comparing the axisymmetric and
non-axisymmetric case we observe that the difference in the Lense-Thirring rotation term results basically
from the fact that -- besides the mass aspect -- the rotation term in the axisymmetric case contains
just the $3$-momentum aspect $p^1$ of the geometry, while the in the non-axisymmetric case
the complete 3-momentum aspect $(p^1,p^2,p^3)$ is present. This clarify the extension of the
Lense-Thirring rotation term for the general boosted case. We note that the momentum
aspects $p^2$ and $p^3$ break the rotational symmetry about the $z$-axis so that the
angular momentum varies as the black hole rotates about the $z$-axis;
$\partial/\partial \phi$ is obviously not a Killing vector of the metric. In the limit of
$p^2$ and $p^3$ going continuously to zero (\ref{eqn22SS}) tends continuously to (\ref{eqn22S})
as expected.
\par
The static limit surface and the ergosphere are also examined in the case of the general
boosted Kerr black hole; as in the axisymmetric case the general boost turns the ergosphere
asymmetric in the direction opposite to the boost.
\par
The singularity at $r=0$, $\Sigma(\theta,\phi)=0$ corresponds to closed curves contained in $2$-dim
planes orthogonal to the direction of the boost determined by $(n_1,n_2,n_3)$. For increasing values
of the boost parameter $|b|$ the radius of the singularity lines initially increase and then decreases.
In the limits $b=0$ (that is, when $b/a=v/c=0$) and $b= \rightarrow \infty$ (that is, when $b/a=v/c \rightarrow 1$)
the closed curves reduces to a point. The point singularities occurring at the equatorial plane
and at the north/south poles (corresponding respectively to $v=0$ and  $v \rightarrow 1$) can be developed
in the Minkowski background to the topology of a ring.
\par Actually the boosted black hole solution can be a natural set for astrophysical processes
connected to the asymmetry of the ergosphere and to electromagnetic dynamical effects that may result from
the rotating black hole moving at relativistic speeds in a direction not coinciding with
the rotation axis. In this setting electromagnetic losses due to translational and rotational motion of the
black hole are expected to occur. These effects may correspond to
the electromagnetic counterpart of the gravitational wave emission by the black hole
having possibly the same order of magnitude,
and can eventually turn out to be important for the astrophysics of highly
energetic bounded sources observed in our actual universe (as for instance active galactic nucleus AGNs)
as we comment below.
\par We envisage that these processes can have applications in modeling the astrophysics
of electromagnetic outflows involving the boost and rotation encompassed in the black hole (\ref{eqn22}).
In fact rotating black holes in electro-vacuum or in a tenuous plasma can produce strong electromagnetic signals
similar to the magnetospheres of rotating pulsars as in the Blandford-Znajek processes\cite{blandford,koide,rezzolla1}.
Another aspect has to do with the motion of the black hole at relativistic speeds in such an environment.
The electromagnetic fields can either be of external origin or due to the motions of the constituents
of the plasma itself.
Electric currents flowing in the plasma may induce a time dependent magnetic field ${\bf B}$
in a plane orthogonal to the rotation axis. Furthermore, since the rotation axis
makes an angle $\theta=\arccos(n_1)$ with the boost direction, a further nonzero electric field
component proportional to $({\bf v}\wedge {\bf B})$ will be present. This makes possible the
appearance of electromagnetic flows that may appear as an electromagnetic counterpart
of late time emissions in the merger of black holes \cite{lyutikov,rezzolla0,rezzolla00,rezzolla11}.
We recall that this boost is inherited from the net momentum flux of the gravitational waves emitted
in the collision and merger of two nonequal mass black holes that generated the remnant.
\par Finally we argue that the application of such mechanisms as engines of relativistic electromagnetic jets from
quasars, pulsars and active galactic nuclei (AGNs) could be properly considered and implemented
in the neighborhood of the general boosted Kerr black hole (\ref{eqn22}).
We are presently examining numerically solutions of Maxwell equations in the
background of the black hole (\ref{eqn22})
taking into account a tenuous plasma as source, with view to the evaluation of the electromagnetic
power emitted in these configurations (cf. also Abdujabbarov et al.\cite{abdu} for the axisymmetric case\cite{ivanoGRG}).
Recently Benavides-Gallego et al.\cite{bambi}
studied weak gravitational lensing around a boosted Kerr black hole\cite{ivanoGRG} in the presence of plasma.

\section*{Acknowledgements}

The author acknowledges the partial financial support of
CNPq/MCTI-Brazil, through the Research Grant No. 308728/2017-3.
The author is also grateful to an anonymous referee for criticisms and comments
that contributed to a substantial improvement of the paper.


\begin{thebibliography}{99}
%
\bibitem{kerr0} R. P. Kerr, Phys. Rev. Lett. {\bf 11}, 237-238 (1963).

\bibitem{schwarz} K. Schwarzschild, Berliner Sitzungsbesichte (Phys. Math. Classe), 189-196 (1916).

\bibitem{gw1} B. P. Abbott et al. (LIGO Collaboration and Virgo Collaboration), Phys. Rev. Lett. {\bf 116}, 061102 (2016).

\bibitem{gw2} B. P. Abbott et al. (LIGO Collaboration and Virgo Collaboration), Phys. Rev. Lett. {\bf 116}, 241103 (2016).

\bibitem{gw3} B. P. Abbott et al. (LIGO Collaboration and Virgo Collaboration), Phys. Rev. Lett. {\bf 119}, 141101 (2017).

\bibitem{gw4} B. P. Abbott et al. (LIGO Collaboration and Virgo Collaboration), Phys. Rev. Lett. {\bf 118} 221101 (2017).

\bibitem{bondi1} H. Bondi, M. G. J. van der Berg, and A. W. K. Metzner, Proc. R. Soc. Lond. A{\bf 269}, 21 (1962).

\bibitem{sachs} R. K. Sachs, Proc. R. Soc. A{\bf 270}, 103 {1962}.

\bibitem{sachs1} R. K. Sachs, J. Math. Phys. {\bf 3}, 908 (1962).

\bibitem{newman} E. T. Newman and R. Penrose, J. Math. Phys. {\bf 7}, 863 (1966).

\bibitem{sachs2} R. K. Sachs, Phys. Rev. {\bf 128}, 2851 (1962).

\bibitem{ivanoGRG} I. D. Soares, Gen. Rel. Grav. {\bf 149}, 77 (2017).

\bibitem{kramer} H. Stephani, D. Kramer, M. A. H. MacCallum, C. Hoenselaers, and E. Herlt,
{\it Exact Solutions of Einstein's Field Equations} (second edition) (Cambridge University Press, Cambridge, 2003).

\bibitem{penrose} E. T. Newman and R. Penrose, J. Math. Phys. {\bf 7}, 863 (1966).

\bibitem{lense} H. Thirring and J. Lense, Phys. Z. 156 (Section 3.10), (1918).

\bibitem{janis} E. T. Newman and A. I. Janis, J. Math. Phys. {\bf 6}, 915 (1965).

\bibitem{poisson} E. Poisson, {\it A Relativist's Toolkit: The Mathematics of Black-Hole Mechanics}
(Cambridge University Press, Cambridge, 2004).

\bibitem{penrose1} R. Penrose, Il Nuovo Cimento {\bf 1} (special number), 252 (1969).

\bibitem{mtw} C. W. Misner, K. S. Thorne, and J. A. Wheeler, {\it Gravitation} (W. H. Freeman and Company, San Francisco, 1970).

\bibitem{mattvisser} M. Visser {\it arXiv: 0706.0622v3} [gr-qc] (2008).

\bibitem{smarr} L. Smarr, Phys. Rev. D{\bf 7}, 289 (1973).

\bibitem{ligo1} B. P. Abbott et al. (LIGO Scientific Collaboration), Rept. Prog. Phys. \textbf{72}, 076901 (2009);
G. M. Harry (LIGO Scientific Collaboration), Class. Quant. Grav. \textbf{27}, 084006 (2010).

\bibitem{blandford} R. D. Blanford, and R. Znajek, Month. Not. Roy. Astron. Soc. {\bf 179}, 433 (1977).

\bibitem{koide} S. Koide, K. Shibata, T. Kudoh, D. L. Maier, Science {\bf 295}, 1688 (2002); S. Koide and T. Baba, Astrophys. J. {\bf 792} (2014).

\bibitem{rezzolla1} P. Moesta, C. Palenzuela, L. Rezzolla, L. Lehner, S. Yoshida, and D. Pollney, Phys. Rev. D{\bf 81}, 064017 (2010).

\bibitem{lyutikov} M. Lyutikov, Phys. Rev. D{\bf 83}, 064001 (2011).

\bibitem{rezzolla0} P. Moesta, D. Alic, L. Rezzolla, O. Zanotti, and  C. Palenzuela,, Astrophys. J. Lett. D{\bf 749}, L32 (2012).

\bibitem{rezzolla00}  D. Alic, P. Moesta, L. Rezzolla, O. Zanotti, and J. L. Jaramillo, Astrophys. J. D{\bf 754}, 36 (2012).

\bibitem{rezzolla11} V. S. Morozova, L. Rezzolla, and B. J. Ahmedov, Phys. Rev. D{\bf 89}, 104030 (2014).

\bibitem{abdu} A. Abdujjabarov, N. Dadhich, and B. Ahmedov, arXiv:1810.08066[gr-qc] (2018).

\bibitem{bambi} C. A. Benavides-Gallego, A. A. Abdujabbarov and C. Bambi, Eur. Phys. J. C (2018) 78:694.


\end{thebibliography}
\end{document}